\documentclass[aps.pra,longbibliography,onecolumn,showpacs,preprintnumbers,amsmath,amssymb]{revtex4-1}
\usepackage{mathrsfs}
\usepackage{natbib}
\usepackage{graphicx}
\usepackage{amsmath}
\usepackage{ulem}
\usepackage{color}
\usepackage[cp1251]{inputenc}
\usepackage[english]{babel}
\usepackage{amsthm,amsfonts,amsmath,amssymb,amscd}

\begin{document}

\newcommand{\bra}[1]{\langle#1|}
\newcommand{\ket}[1]{|#1\rangle}
\newcommand{\op}[2]{\hat{\textbf{#1}}_{#2}}
\newcommand{\dagop}[2]{\hat{\textbf{#1}}_{#2}^\dag}
\newcommand{\zhukov}[1]{{\color{blue}{#1}}}
\newcommand{\remove}[1]{{\color{red}{#1}}}

\title{Parametrically driven hybrid qubits-photon systems: dissipation-induced quantum entanglement and photon production from vacuum}
\author{S. V. Remizov$^{1,2}$, A. A. Zhukov$^{1,3}$, D. S. Shapiro$^{1,2}$, W. V. Pogosov$^{1,4}$, Yu. E. Lozovik$^{1,5,6}$}

\affiliation{$^1$N. L. Dukhov All-Russia Research Institute of Automatics, 127055 Moscow, Russia}
\affiliation{$^2$V. A. Kotel'nikov Institute of Radio Engineering and Electronics, Russian Academy of Sciences, 125009 Moscow, Russia}
\affiliation{$^3$National Research Nuclear University (MEPhI), 115409 Moscow, Russia}
\affiliation{$^4$Institute for Theoretical and Applied Electrodynamics, Russian Academy of
Sciences, 125412 Moscow, Russia}
\affiliation{$^5$Institute of Spectroscopy, Russian Academy of Sciences, 142190 Moscow region,
Troitsk, Russia}
\affiliation{$^6$Moscow Institute of Physics and Technology, 141700 Moscow region, Dolgoprudny, Russia}

\begin{abstract}
We consider a dissipative evolution of parametrically-driven qubits-cavity system
under the periodical modulation of coupling energy between two subsystems,
which leads to the amplification of counterrotating processes.
We reveal a very rich dynamical behavior of this hybrid system. In particular, we find that the energy dissipation in one of the subsystems can enhance quantum effects in another subsystem. For instance, optimal cavity decay assists to stabilize entanglement and quantum correlations between qubits even in the steady state and to compensate finite qubit relaxation. On the contrary, energy dissipation in qubit subsystem results in the enhanced photon production from vacuum for strong modulation, but destroys both quantum concurrence and quantum mutual information between qubits. Our results provide deeper insights to nonstationary cavity quantum electrodynamics in context of quantum information processing and might be of importance for dissipative quantum state engineering.
\end{abstract}

\author{}
\maketitle
\date{\today }

\section{Introduction}

Hybrid light-matter systems are considered as promising candidates for the implementation of quantum information and computation devices  \cite{DiCarlo, MSS, Nation,  Clarke, Schoelkopf, Ashhab, Reuther}. There are various physical platforms which can be used to organize an appropriate light-matter interaction \cite{Macha,Astafiev,Rakhmanov, Fistul,ZKF,SMRU,remizov,Brandes,Zou}. Some of the realizations are based on macroscopic or mesoscopic quantum systems such as superconducting Josephson circuits \cite{MSS, Orlando,mooij} or quantum dots coupled to microcavities \cite{Morton, Schuster, Reithmaier}. Other realizations utilize more microscopic systems such as NV-centers in diamond \cite{nv-centers-0,nv-centers, nv-centers-1} or trapped ions \cite{Blatt}.

Macroscopic quantum systems and especially superconducting Josephson qubits coupled to microwave transmission-line cavities are characterized by a high degree of tunability and flexibility -- various effective parameters of a hybrid system can be made even {\it dynamically} tunable {\it in-situ} \cite{Liao, tunable1, tunable2, tunable3}. This opportunity allows to implement parametrical processes, which can hardly be activated using natural quantum systems. For example, several years ago a first observation of one of the most intriguing nonstationary quantum electrodynamical (QED) phenomena known as the dynamical Casimir effect was reported in superconducting systems \cite{DCE1, DCE2,Nation}. This was achieved by embedding an additional SQUID at the end of a coplanar waveguide and changing the magnetic flux through it, which is equivalent to the modulation of the boundary condition for the electromagnetic field.
Recently, we suggested a realization of new nonstationary QED effect, the dynamical Lamb effect \cite{Lozovik1}, in superconducting systems \cite{paper1,paper2,paper3}. It can be considered as a parametric excitation of qubit (atom) due to the nonadiabatical modulation of its Lamb shift or shaking of photon dressing. A natural way to observe it is to use hybrid light-matter systems with the dynamically tunable coupling between two subsystems.

For the construction of quantum computation and simulation devises, it is necessary to have at
least few qubits, the entanglement between them being a key resource \cite{Amico}. Initially it was believed that the coupling to the environment, usually referred to as decoherence, always destroys this resource. However, in reality the situation is far more complex. Indeed, it was recognized some time ago \cite{Verstraete2008} that a special design of the interactions with the environment can be utilized to implement universal quantum computation. In particular, reservoir engineering can be used for generating steady-state entanglement at finite temperatures \cite{Milman}. Having in a mind a possibility to activate various QED effects in tunable artificial quantum systems, it is of interest to explore the dynamics of entanglement under parametric excitations of different types.

In this paper, we consider parametric process which is generated by the periodic
modulation of the qubit-cavity coupling and in presence of energy dissipation. This parametric driving
is able to greatly enhance certain quantum effects, such as parametric qubit excitation via counterrotating
wave terms of the Hamiltonian bringing weakly-interacting system effectively to the regime of strong
light-matter coupling \cite{paper1,paper2,Nori2016,Liberato}.
We here show that a periodic modulation of a qubit-cavity couplings is able to generate quite high entanglement of the two qubits even in the limit of a weak interaction. Moreover, depending on the system parameters and the modulation signal, the entanglement can survive in the steady-state. Surprisingly, finite and optimal cavity decay rate assists generation of the steady-state entanglement of qubits.

This paper is organized as follows. In Section II, we describe the system under consideration and
outline our theoretical model. In Section III, we present results of our calculations for the decoherence-free systems. In Section IV we analyze the effect of decoherence. We conclude in Section V.

\section{Hamiltonian and basic equations}

We consider a system of two identical qubits in a circuit QED architecture. These qubits are coupled to a single-mode quantum resonator. Qubits interact with each other only through a photon degree of freedom. The Hamiltonian of a hybrid qubits-resonator system can be represented as
\begin{equation}
H(t)=\omega a^{\dagger }a + \sum_{j=1,2} \epsilon _{j} \sigma_{j,+}\sigma_{j,-} + \sum_{j=1,2} G_{j}(t) (\sigma_{j,+}+\sigma_{j,-})(a^{\dagger }+a),
\label{Hamiltonian}
\end{equation}
where $j$ labels qubits, $a^{\dagger }$ and $a$ are secondary quantized photon creation and annihilation operators,
$\sigma_{j,+}$, $\sigma_{j,-}$ are Pauli operators related to qubit degrees of freedom, while $G_j (t)$ is the interaction constant between a given qubit and photon field. We assume that constants  $G_j (t)$ are dynamically tunable. Such a modulation can be achieved using different schemes and, in principle, various physical platforms. For example, it is possible to use flux qubits with an additional SQIUD or two strongly coupled charge qubits (transmons) based on which a single effective two-level system can be created, see, e.g., Refs. \cite{tunable1,tunable2}. Another approach is based on a three-level superconducting system (transmon) under the coherent drive \cite{tunable3}. We hereafter assume that $G_1(t)=G_2(t)=G(t)$, so that coupling between qubit and photon subsystems is controlled by a single parameter. The last term in the right-hand side of Eq. (\ref{Hamiltonian}), which describes the interaction between subsystems, contains both contributions known as rotating wave term, $G(t) V_1= G(t) \sum_{j} (\sigma_{j,+}a+\sigma_{j,-}a^{\dagger })$, and counterrotating wave term, $G(t) V_2= G(t) \sum_{j} (\sigma_{j,+}a^{\dagger }+\sigma_{j,-}a)$.

In order to take into account decoherence effects on simplest grounds, we address dissipative evolution by solving numerically the master equation
\begin{equation}
\partial_t\rho(t)-\Gamma[\rho(t)]=-i[H(t),\rho(t)],
\label{Lindblad}
\end{equation}
where $\rho(t)$ is a density matrix of qubit and photon subsystems. The matrix $\Gamma[\rho]$ depends on rates of energy dissipation in the cavity $\kappa$, in each of the two identical qubits $\gamma$, as well as on pure dephasing rate $\gamma_{\varphi}$. It is given by $
\Gamma[\rho]=\kappa(a\rho a^{\dagger } -\{ a^{\dagger }a, \rho\}/2)+\sum_{j=1,2}\left(\gamma(\sigma_{j,-}\rho\sigma_{j,+} - \{\sigma_{j,+}\sigma_{j,-},\rho\}/2)+ \gamma_{\varphi}(\sigma_{j,z}\rho\sigma_{j,z}-\rho)\right)$.
Relations between  $\kappa$ and $\gamma$ can be very different for various realizations of hybrid qubit-photon systems. For instance, in the state-of-art superconducting qubits coupled to microwave resonators the pure decoherence rate $\gamma_{\varphi}$ is typically
of the same order as relaxation $\gamma$, while both these quantities are significantly larger than the relaxation rate in a cavity, $\gamma\gg \kappa$. For NV-centers, as well as for other microscopic realizations of hybrid systems, the situation is just the opposite -- relaxation rate of cavity is typically much larger than decoherence rates of the two-level systems.

In this article, we are mainly focused on such characteristics of qubit subsystem, as quantum concurrence $C$, which is a measure of entanglement between qubits, and quantum mutual information $I$, which is a measure of correlations between them. Since our system consists of two qubits and photon subsystem, the latter degree of freedom has to be somehow eliminated in order to quantify an entanglement of qubits based on two-qubit density matrix $\rho_{qq}$ of size $4  \times 4$.  We perform such an elimination in two ways. According to the first algorithm, we simply trace out photons $\rho_{qq}  =  \mathop{\rm Tr}_{\rm ph} \rho$ and then find a concurrence of qubits $C$ based on the obtained $\rho_{qq}$. Within the second approach, we treat $\rho_{qq}$ as a block of size $4  \times 4$ of full density matrix $\rho$ related only to states with fixed and given number of photons $i$. We then find a 'conditional concurrence' $C_i$ corresponding to given $i$. In both cases, $C$ or $C_i$ is evaluated from $\rho_{qq}$ in a standard way being defined as $\tilde{C} = \sqrt{\lambda_1} - \sqrt{\lambda_2} - \sqrt{\lambda_3} - \sqrt{\lambda_4}$ provided $\tilde{C}$ is positive and zero otherwise, where $\lambda_{i}$ are sorted in descending order eigenvalues of matrix $\rho_{qq} \sigma \rho_{qq}^* \sigma$,  $\sigma = \sigma_y \otimes \sigma_y$.

For the von Neumann entropy we also use a standard definition $S = - \mathop{\rm Tr} \rho \ln \rho$. An important quantity in the case of a bipartite system is quantum mutual information, $I$. It is defined as $I=S_1+S_2 - S_{12}$, where $S_{12}$ is entropy of a whole system, while $S_{1}$ and $S_{2}$ stand for two subsystems. The entropy $S_{1}$ ($S_{2}$) is calculated by tracing out degrees of freedom for subsystem 2 (subsystem 1). We use a similar definition for the quantum mutual information between two qubits in presence of a photonic subsystem. Photonic degree of freedom thus must be also traced out when finding $S_{12}$, $S_{1}$, and $S_{2}$.

We assume that in the initial moment qubits are uncoupled from the resonator, so both qubit and photon subsystems are in their ground states, $|\downarrow \downarrow, 0\rangle$. At certain moment, interaction constant $G(t)$ starts to be modulated periodically in time. Under these conditions, counterrotating terms responsible for the dynamical Lamb effect \cite{paper1}, are essential to impose a dynamics of a hybrid system, since rotating wave terms are unable to change the excitation number.

For a single qubit coupled to the resonator, strongly enhanced occupation of qubit excited state is achieved, when qubit is in a resonance with the cavity and when the modulation frequency of coupling constant is twice the cavity frequency, which dramatically enhances a transition from $|\downarrow, 0\rangle$ to $|\uparrow, 1\rangle$ bringing the occupation of $|\uparrow, 1\rangle$ to large values of the order of 1 even if $|G(t)| \ll \omega$ \cite{paper1,paper2}. It is expected that for a two-qubit system the behavior of both qubits occupations will be qualitatively similar. Theoretical treatment of this particular case can be simplified by using a separation over fast and slow degrees of freedom and performing time averaging. Physically, it corresponds to the situation when there are two types of oscillations in the system -- small-amplitude fast oscillations with the frequency of the qubit excitation and large-amplitude oscillations with frequencies of the order of the Rabi frequency. These fast oscillations are eliminated by a time averaging procedure \cite{paper1,paper2}.

Under these conditions and in the limit $|G(t)| \ll \omega$, system's dynamics is governed by two Fourier components \cite{paper1,paper2} defined as
\begin{eqnarray}
\left\langle G(t)\right\rangle_{t} \equiv p , \nonumber \\
\left\langle G(t)\exp (-2i\omega t)\right\rangle_{t} \equiv  q,
\label{pq}
\end{eqnarray}
where averaging is performed with respect to the time period $\pi / \omega$, while it is also assumed that $G(t)$ is symmetrical with respect to the time reversal. It is convenient to recast $p$ and $q$ as $p=g \theta$, $q=g (1-\theta)$, where $0 \leq \theta  \leq 1$. The parameter $p$ provides a strength of the interaction in the Tavis-Cummings channel, which conserves excitation number, while $q$ is responsible for the counterrotating processes changing excitation number by $\pm 2$. Note that $\theta < 0.5$ corresponds to sign-alternating $G(t)$ oscillations. Let us stress that we only focus on the case $G(t)=G(-t)$.

Thus, we consider a Hamiltonian in the interaction picture, the coupling between the qubit and photon subsystems being
\begin{equation}
g \sum_{j=1,2} [\theta (\sigma_{j,+}a+\sigma_{j,-}a^{\dagger }) + (1-\theta)(\sigma_{j,+}a^{\dagger }+\sigma_{j,-}a)],
\label{Hamiltonian1}
\end{equation}
which is not dependent explicitly on time. We then analyze system dynamics after the sudden switching of the coupling constant from zero to $g$, the initial state of the system being $|\downarrow \downarrow, 0\rangle$.

\section{Dynamics of decoherence-free system}

Before considering the dynamics of the system with energy dissipation, it is reasonable to explore its evolution without decoherence. It can be addressed using a solution of the Schr\"{o}dinger equation numerically. Explicit results can be obtained at special points $\theta=0$, $1/2$, and $1$. The analysis for these three points is presented in Appendix A.

\begin{figure}[h]
\includegraphics[width=0.45\linewidth]{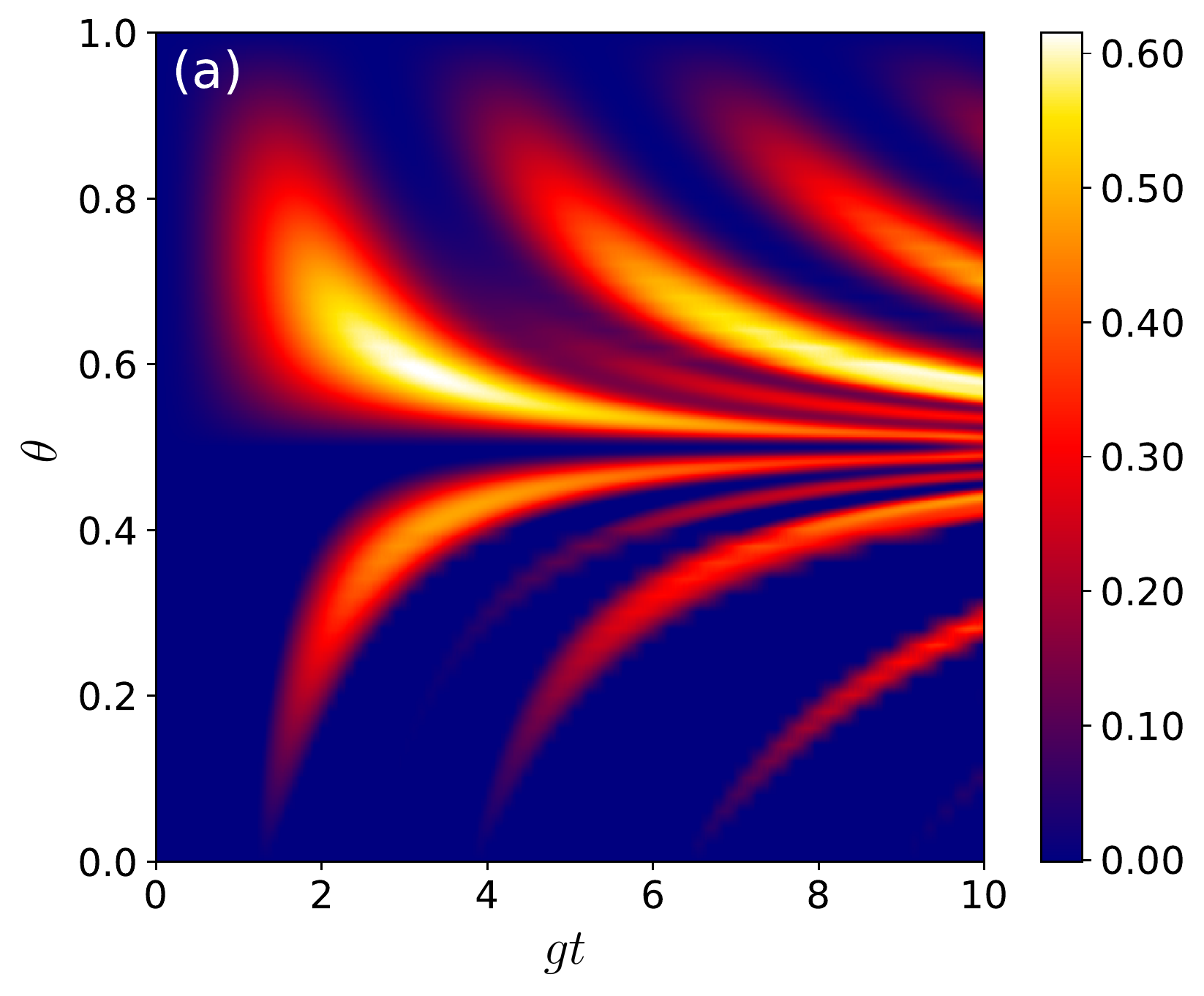}
\includegraphics[width=0.45\linewidth]{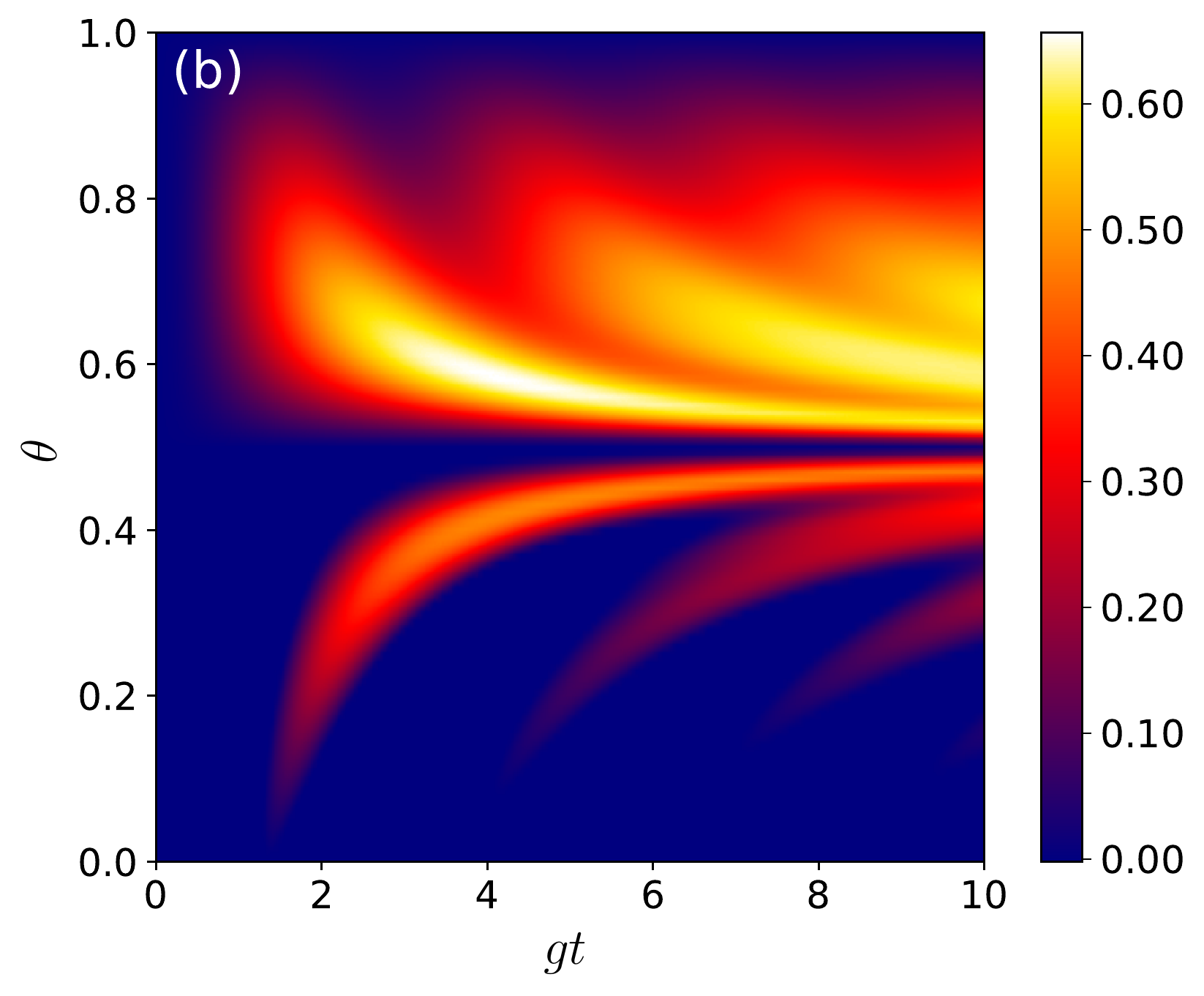}
\caption{
\label{nodecoherence}
Color map for the time evolution of the quantum concurrence between the two qubits at $g_0=0.05 \omega$ and different values of parameter $\theta$ at cavity relaxation rate $\kappa=0$ (a) and $0.01 \omega$ (b).}
\end{figure}

Our results show that the parametric driving leads to the generation of a rather significant quantum concurrence between initially independent qubits. Fig. \ref{nodecoherence} (a) displays the time evolution of the concurrence for different values of $\theta$ at $g=0.05 \omega$. There exist oscillations of this quantity with nearly the frequency of Rabi oscillations. The maximum value of the concurrence is as high as 0.6 despite of the fact that $g$ is a small quantity compared to $\omega$. This is a direct consequence of a resonance between excitation energies of both qubits and cavity frequency, as well as a special choice of the modulation frequency. Note that, for illustration purposes, we hereafter take relatively large $g$ (as well as $\gamma$ and $\kappa$ in the next Section) in order to shorten the transient processes, but qualitative picture remains the same for smaller values of these quantities.

The attractive feature of the quantum concurrence dynamics in Fig. \ref{nodecoherence} is that areas with maximum $C$ are elongated parallel to the time axis, so that in practice it should be easier to attain these high values.

\begin{figure}[h]
\includegraphics[width=0.45\linewidth]{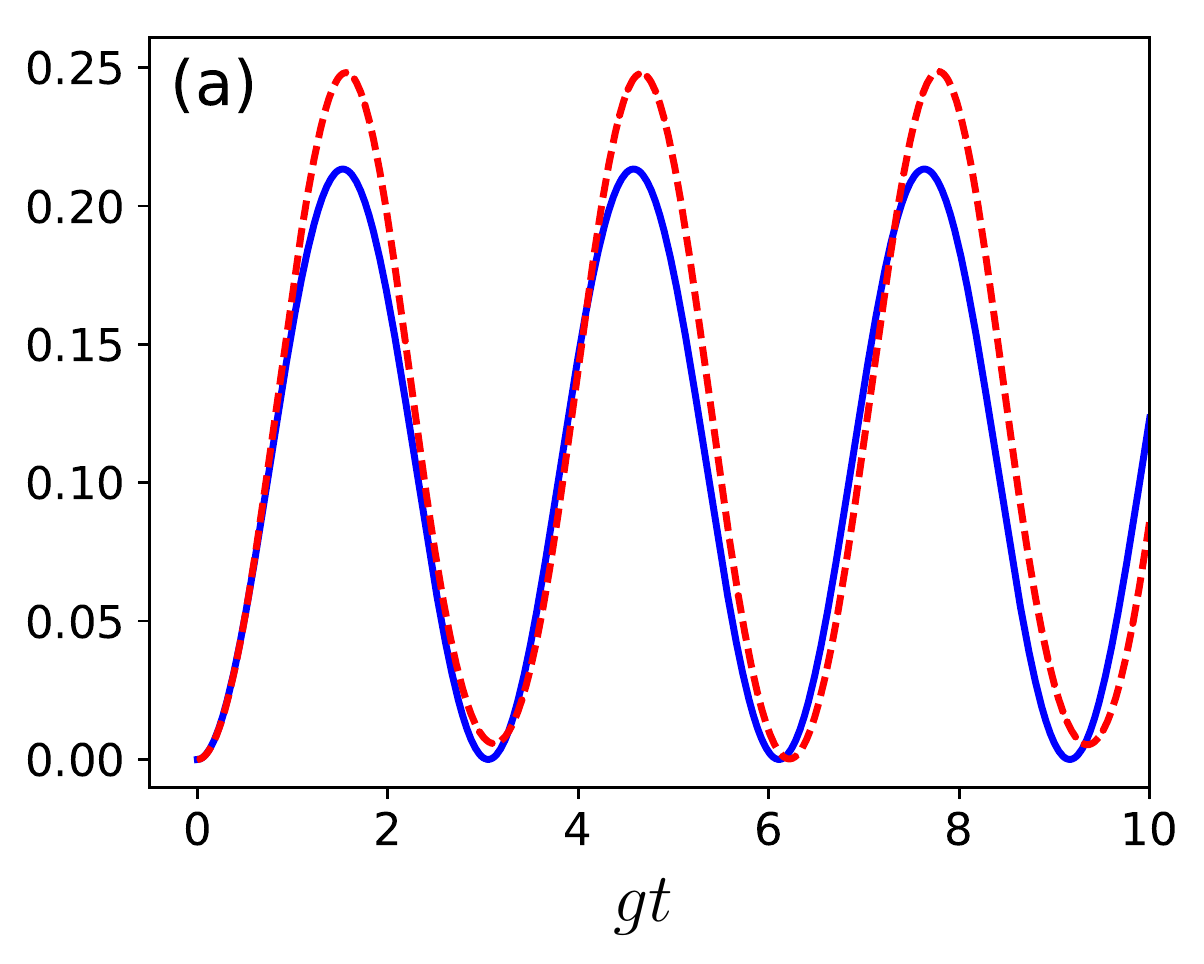}
\includegraphics[width=0.45\linewidth]{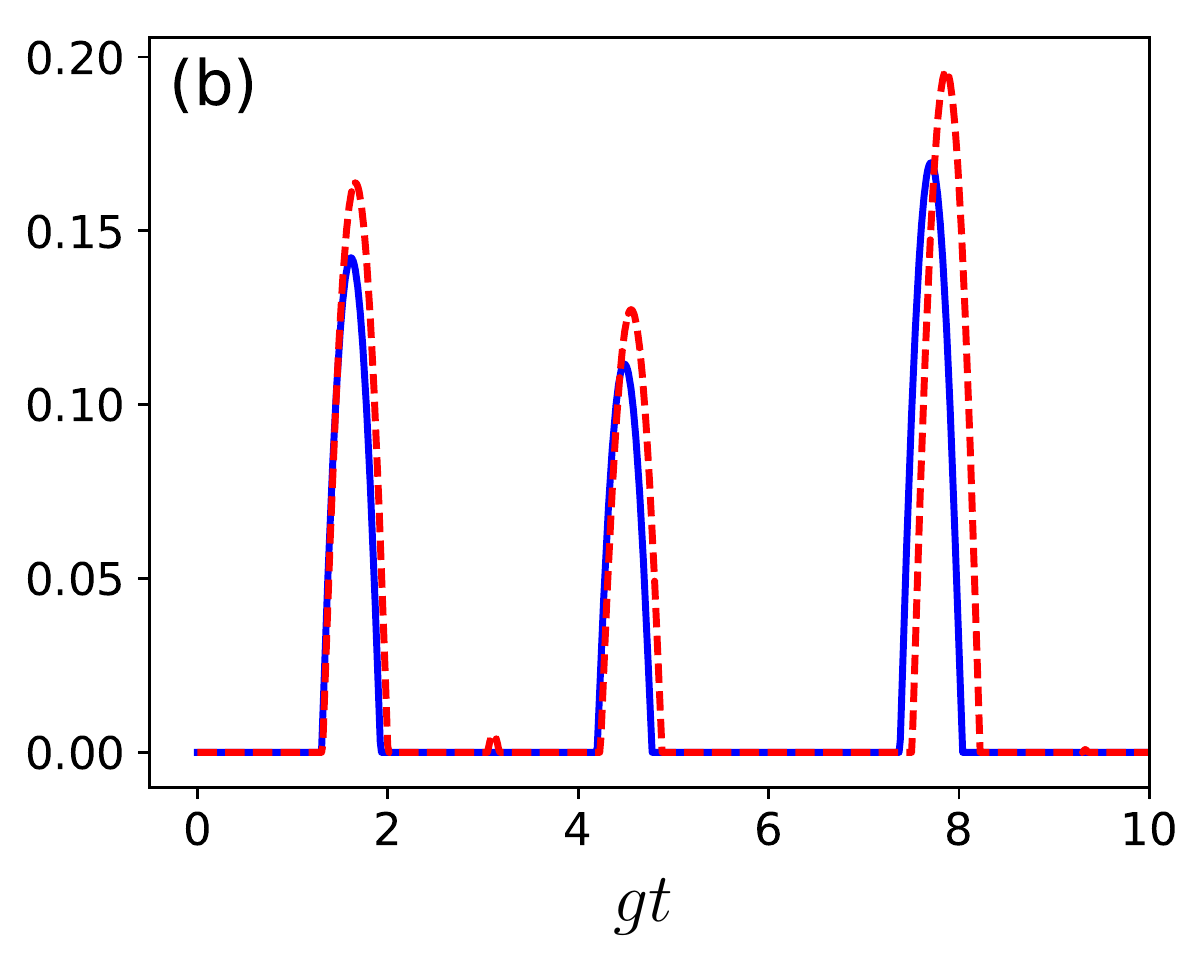}
\caption{
\label{explic}
Time evolution of quantum concurrence between two qubits at $g=0.05 \omega$ for $\theta=0.84$ (a) and $\theta=0.16$ (b) calculated analytically (blue solid lines) using perturbation theory around Tavis-Cummings (a) and Anti-Tavis-Cummings (b) regimes and numerically (red dashed lines).}
\end{figure}

In Fig. \ref{explic}, we show the dynamics of quantum concurrence found numerically and analytically using the results of Appendix A. Namely, at large $\theta$ (a), a perturbation theory around Tavis-Cummings regime is constructed, while at $\theta$ close to 0 (b) a perturbation expansion around Anti-Tavis-Cummings regime is used. There exists a good agreement between the results of numerical and analytical approaches. Note that the agreement, of course, becomes better and better when approaching points $\theta=0$ and 1. Near the Tavis-Cummings regime $C$ is nonzero at any time instance, while in the vicinity of Anti-Tavis-Cummings regime it is finite only within some time intervals. The widths of these intervals vanish as $\theta \rightarrow 0$. As shown in Appendix A, the nonzero concurrence is provided by a superposition of qubit states $\ket{\downarrow \downarrow}$ and $\ket{\uparrow \uparrow}$. This is confirmed by numerical results in the whole domain  $\theta \in [0,1]$.

\section{Influence of energy dissipation}

Let us begin our analysis with the situation when there is only cavity relaxation in the system.
In the opposite situation, when there is only qubit relaxation, this decoherence channel
along easily suppresses quantum concurrence between the two qubits, so that no entanglement survives at $t \rightarrow \infty$. These two situations correspond to
systems which are characterized by the separation of time scales for energy dissipation in their photon
and qubit subsystems. Such a separation is indeed common for various physical realizations of qubit-photon hybrid systems.
Thus, by ignoring the energy dissipation in one of the subsystems, we can trace a time evolution within some initial
period until decoherence effects in another subsystem come into play.

\subsection{The effect of cavity relaxation}

We now address the effect of cavity decay. Fig. \ref{nodecoherence} (b) shows time evolution of quantum concurrence  for $\kappa=0.01 \omega$. We see that the oscillations of $C$ persist, but they become smeared as $\kappa$ increases. Again, it can be realized that not so high time precision is needed to attain regions with high quantum concurrence due to the smearing and elongation of patterns along the time axis.

\subsection{The effect of qubit relaxation}

\begin{figure}[h]
\includegraphics[width=0.45\linewidth]{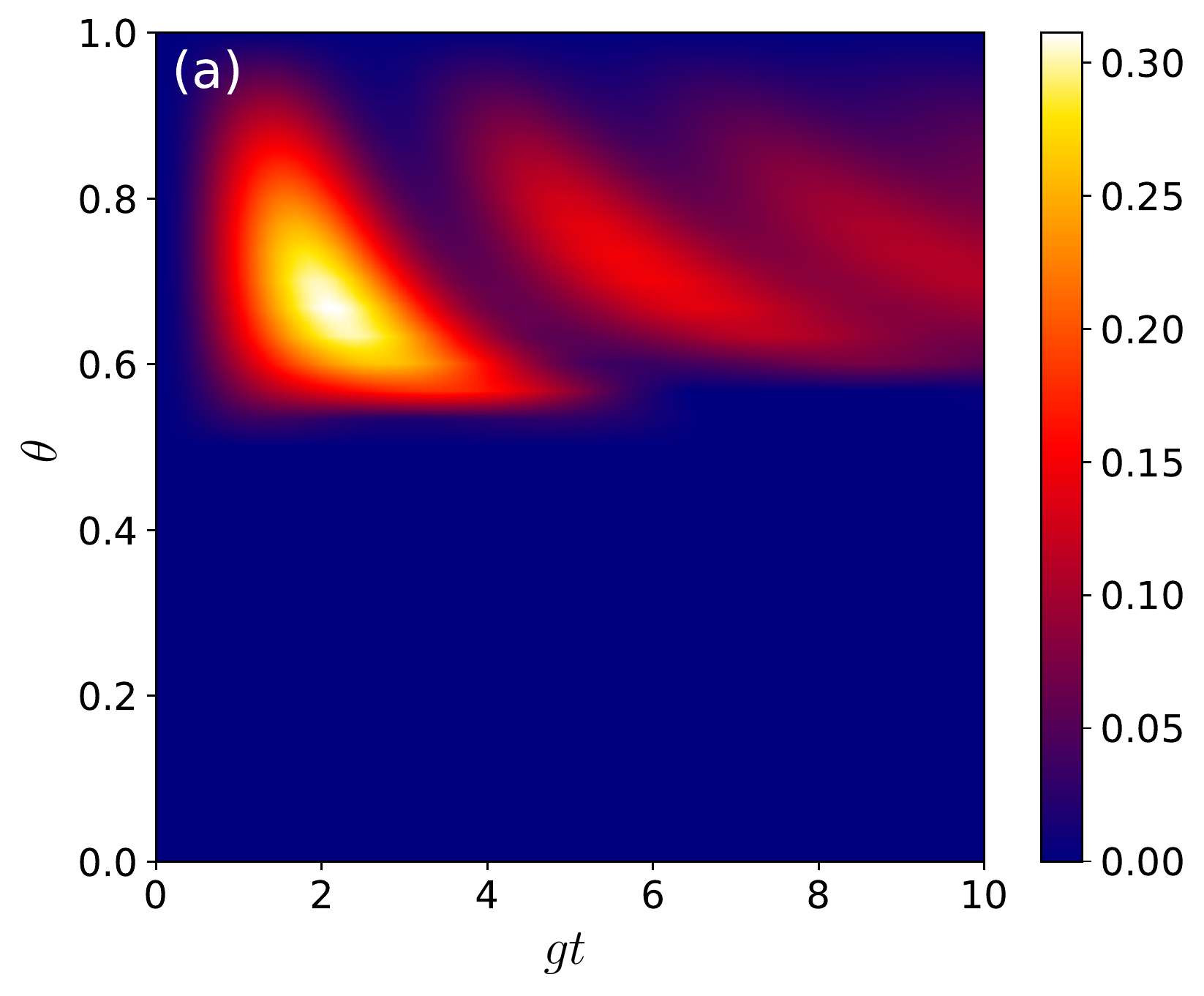}
\includegraphics[width=0.45\linewidth]{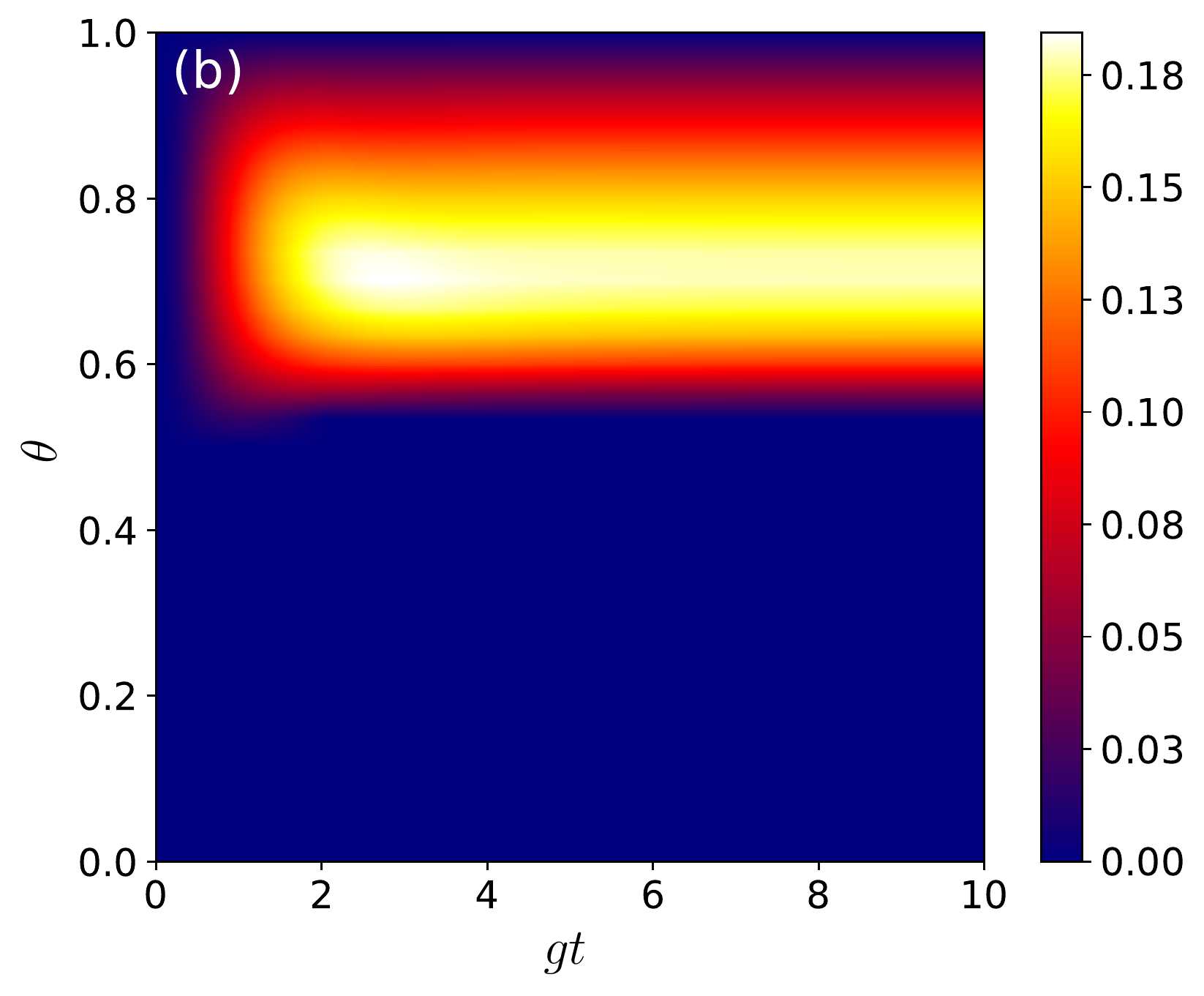}
\caption{
\label{qubit}
Color map for the time evolution of the quantum concurrence between the two qubits at $g=0.05$ and different values of parameter $\theta$ at  $\kappa=0.005 \omega$, $\gamma=0.02 \omega$ (a) and $\kappa=0.05 \omega$, $\gamma=0.01 \omega$ (b).}
\end{figure}

Now let us take into account qubit relaxation $\gamma$, which plays a destructive role for the entanglement between qubits. Typical maps for the time evolution of the quantum concurrence are shown in Fig. \ref{qubit}. It is seen from this figure that two distinct scenarios are possible - either the concurrence goes to zero as $t \rightarrow \infty$ or it remains finite. The latter scenario occurs in the domain $\theta>1/2$. In particular, $C$ vanishes within very short time interval at $\theta < 1/2$. Interestingly, nonzero $\gamma$ tends to destroy the concurrence at the steady state, while nonzero $\kappa$ can restore it. This result is intriguing because in this case decoherence assists purely quantum effects to be amplified, which is definitely a counterintuitive result.

Remarkably, the effect of the enhancement of quantum effects in one subsystem due to the increase of decay rate in another subsystem occurs also in the domain $\theta < 1/2$ -- finite relaxation rate in a qubit leads to the amplification of photon generation from vacuum. In this case, we deal with the enhancement of quantum effects in a photon subsystem due to the finite decay rate in qubit subsystem. The explanation is that fully polarized states $|\uparrow \uparrow, n\rangle$ cannot be driven towards larger $n$'s by $V_2$, since for this at least one of the qubits must be in its ground state. Thus, some upward paths in energy ladder become blocked. However, if a finite $\gamma$ is present in the system, one of the qubit can be deexcited, and then excited again via $V_2$ to the state with $n+1$ photons. This leads to the occupation of both subsets of energy levels with even and odd excitation numbers. It turns out that for a strong (sign-alternating) modulation, i.e., when $\theta < 0.5$, such $\gamma$-assisted excitation becomes very efficient and it overcomes Tavis-Cummings processes generated by $V_1$. However, quantum correlations in qubit subsystem in this case become suppressed. Due to the resulting intensive  photon production, the concurrence between the qubits is suppressed, since all possible states of qubit subsystem become populated.

Note that previously we have found a similar effect of enhanced photon generation from vacuum for a single qubit coupled dynamically to the resonator \cite{paper2}. If qubit subsystem consists of $N>1$ qubits, upward paths in energy ladder due to $V_2$ become longer, since they contain $N$ segments corresponding to the full polarization of qubit subsystem. Nevertheless, the role of finite $\gamma$ in the case of general $N$ is similar. Thus, this effect is rather generic and it is linked to the two-level nature of individual qubits.

\subsection{Quantum concurrence and quantum mutual information in the steady state}

In order to more deeply analyze the effect of dissipation-assisted quantum concurrence stabilization at  $\theta>0.5$, let us consider the steady state of the system, i.e., the state at $t\rightarrow\infty$ after the beginning of the parametric driving.

\begin{figure}[h]
\includegraphics[width=0.45\linewidth]{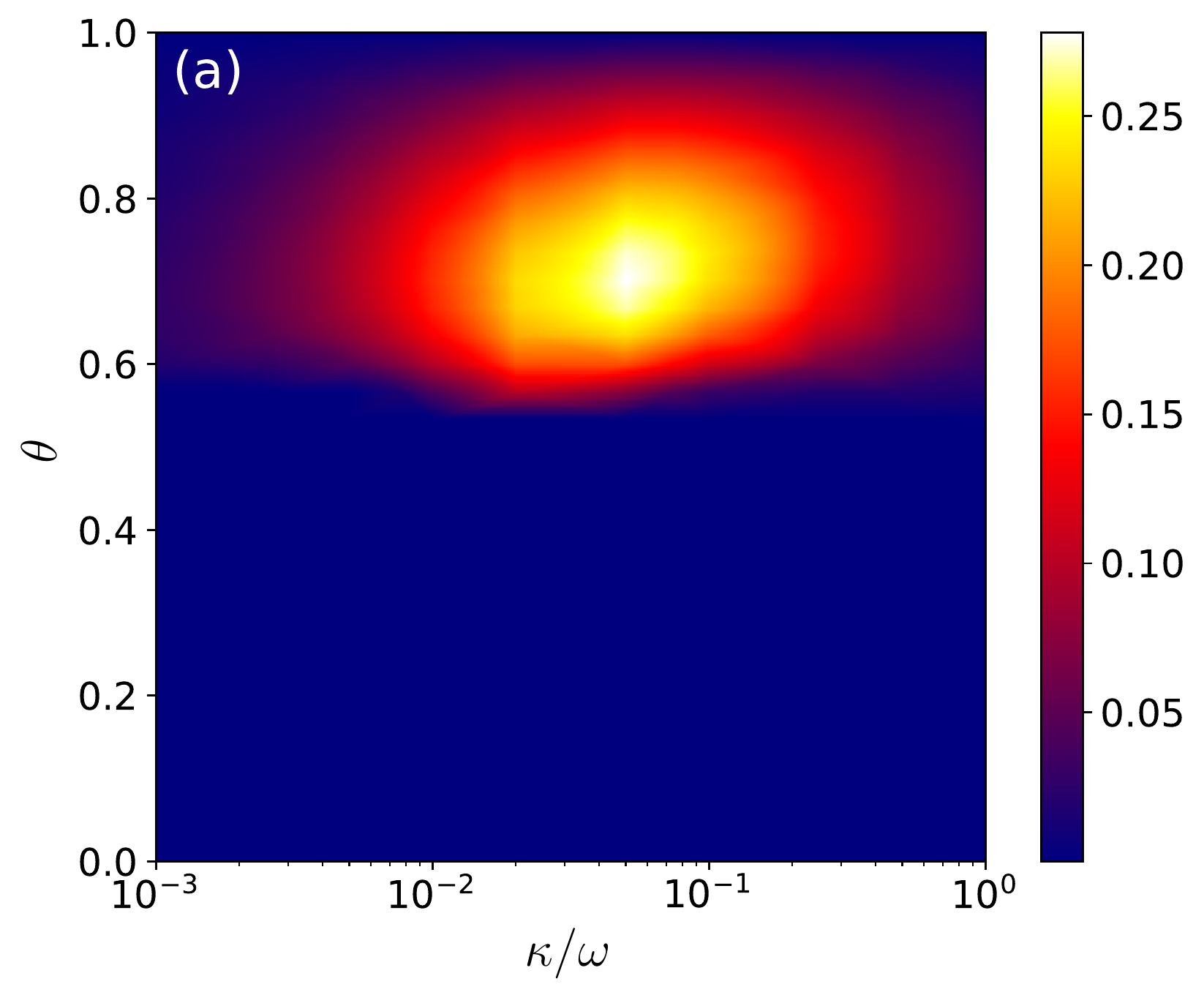}
\includegraphics[width=0.45\linewidth]{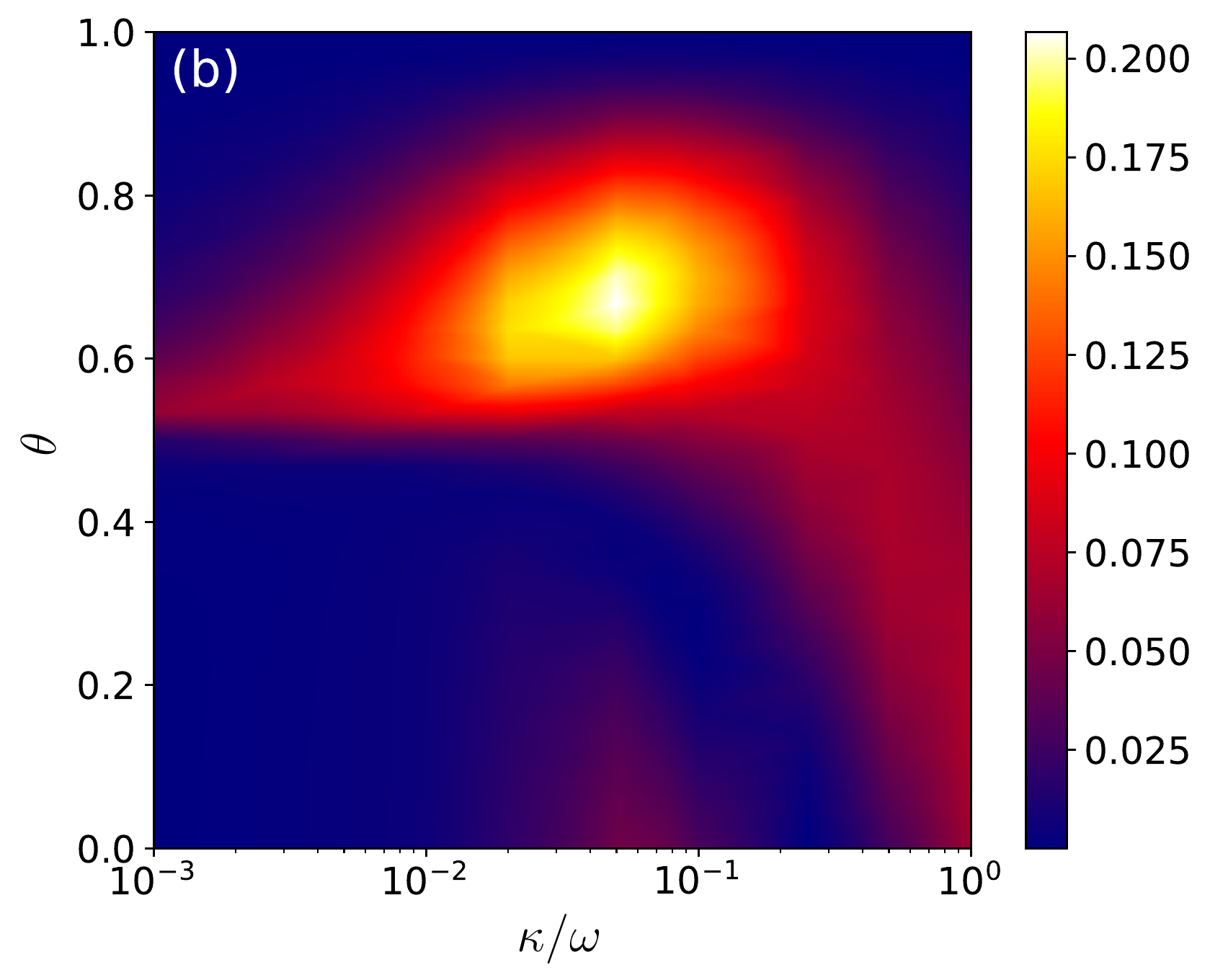}
\caption{
\label{steady}
Color map for the quantum concurrence (a) and mutual information (b) between the two qubits in the steady state at fixed $g=0.05 \omega$ and qubit relaxation rate $\gamma=0.01 \omega$ for different values of $\theta$ and cavity relaxation rate. }
\end{figure}

Figure \ref{steady} (a) shows the map of the concurrence in the steady state in the plane of $\theta$ and $\kappa$ at fixed $\gamma$. The major conclusion is that the quantum concurrence of qubits in a certain domain of parameters survives for an infinitely long time despite of the finite energy dissipation rate both in qubit and cavity subsystems. This is a consequence of a special parametric driving, which allows to keep quantum nature of the system under the consideration 'forever'. It is quite remarkable that highest $C$ and $I$ are attainable at some optimal $\kappa$, which is not too high or too small. Thus, the phenomenon we reveal indeed can be refereed to as the dissipation-induced quantum concurrence.

Figure \ref{steady} (b) shows a similar map of the quantum mutual information. This quantity demonstrates a behavior analogous to the behavior of a concurrence. Quantum mutual information shows that there exist correlations between the two qubits. Notice that there are some regions in the map, where $C$ is zero, while $I$ is nonzero. This implies that qubits in the corresponding parametric region are correlated but not entangled.

\begin{figure}[h]
\includegraphics[width=0.45\linewidth]{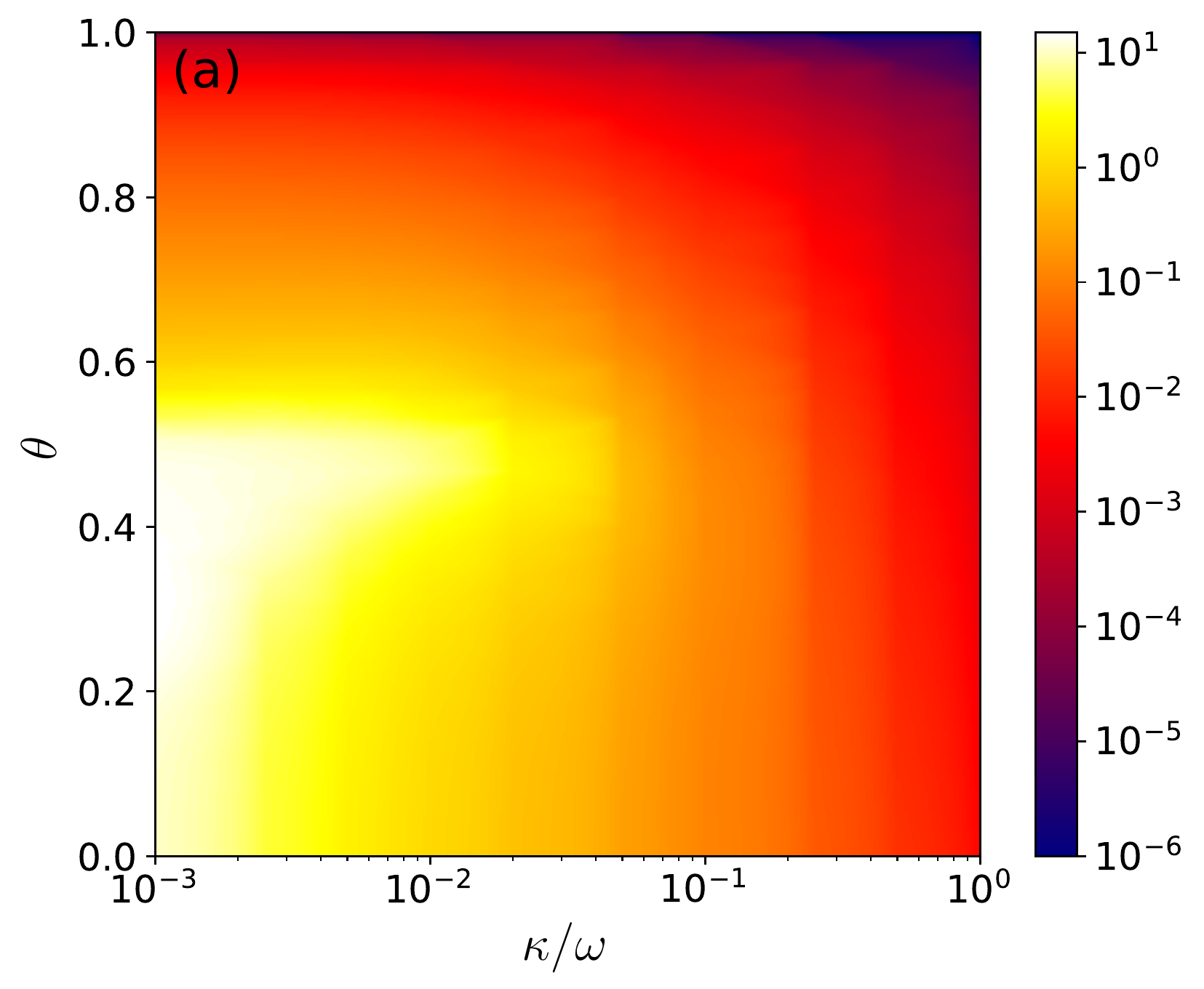}
\includegraphics[width=0.45\linewidth]{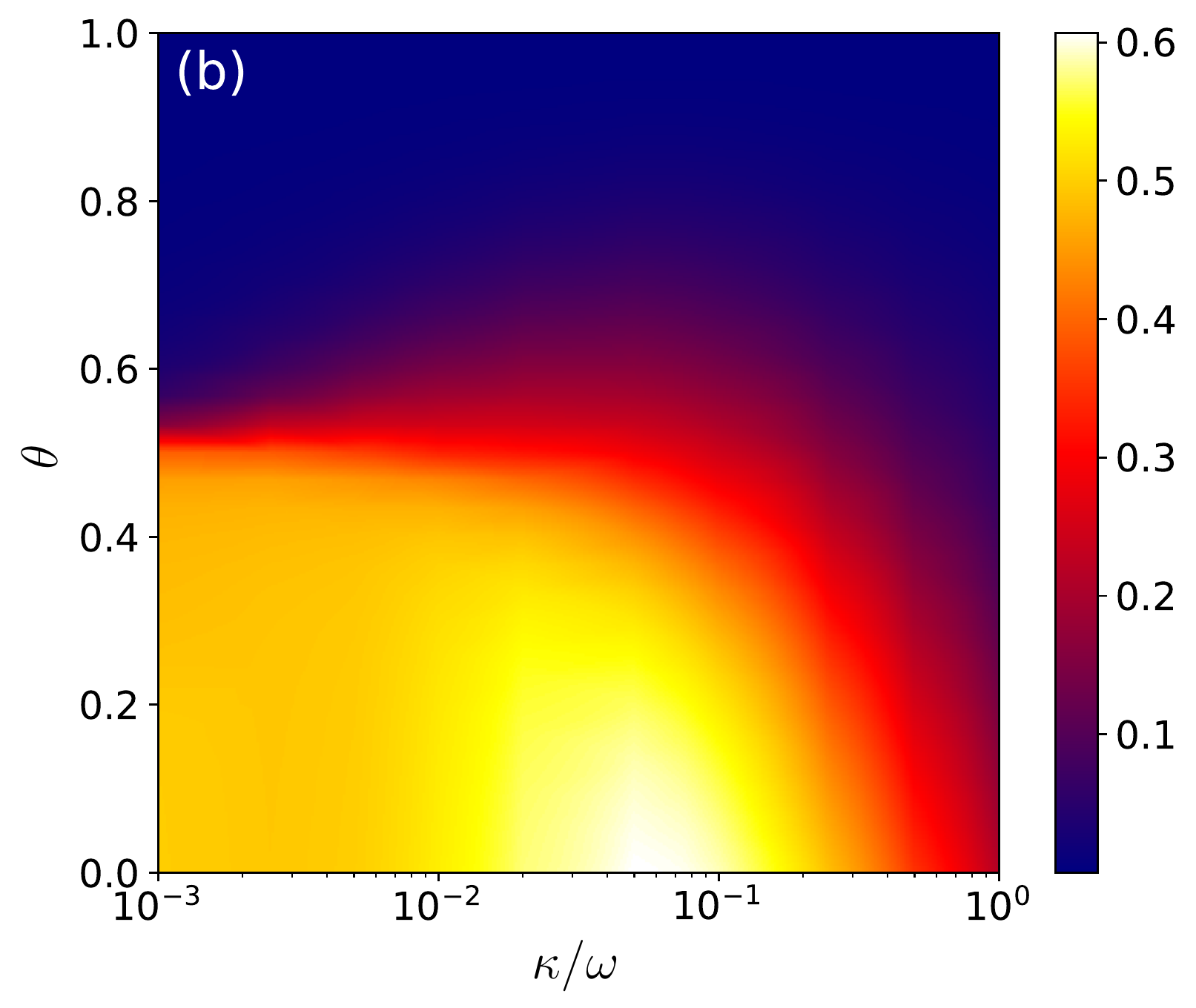}
\caption{
\label{nphnq}
Color map for the mean number of photons (a) and population of the excited state of any of two qubits (b) in the steady state at fixed $g=0.05 \omega$ and qubit relaxation rate $\gamma=0.01 \omega$ for different values of $\theta$ and cavity relaxation rate. }
\end{figure}

In order to see correlations between quantum concurrence and mutual information from one side and mean photon number and qubits excited states occupation from another side, in Fig. \ref{nphnq} we also plot the maps of mean number of photons and population of the excited state of any of the two qubits in the steady state (these populations for two qubits are the same as they are identical). The inspection of Fig. \ref{steady} reveals that if $\theta \lesssim 0.5$ (sign-alternating $G(t)$ oscillations) concurrence is suppressed in the steady state and this is linked to the intensive generation of photons, see Fig. \ref{nphnq} (a) and Ref. \cite{paper2}. In the opposite domain $\theta \gtrsim 0.5$ there is a region with enhanced quantum concurrence and mutual information. It corresponds to some crossover region in maps of Fig. \ref{nphnq}, where both mean number of photons and qubit excited state population start decreasing.

More close examination involving computation of conditional concurrences $C_i$ shows that the finite quantum concurrence between the two qubits is mainly caused by the states with zero photon number. Thus, a qualitative explanation of the effect of dissipation-induced quantum concurrence can be as follows. The analytical derivation, presented in Appendix A for decoherence-free system, shows that at $\theta$ small a nonzero concurrence is generated by the piece of the total wave function, which is a superposition of $\ket{\downarrow \downarrow, 0}$ and $\ket{\uparrow \uparrow, 0}$. A nonzero $\kappa$ allows to create a kind of a circle consisting of three processes.
Namely, the system is excited from the lowest-energy state $|\downarrow \downarrow, 0\rangle$ to the state $|\uparrow \downarrow, 1\rangle+|\downarrow \uparrow, 1\rangle$ via $V_2$. Then, finite occupations of the states $|\uparrow \uparrow, 0\rangle$ and $|\downarrow \downarrow, 2\rangle$ are produced by $V_1$. The first one, $|\uparrow \uparrow, 0\rangle$, gives nonzero $C$ as shown in the Appendix A. However, the state $|\uparrow \downarrow, 1\rangle+|\downarrow \uparrow, 1\rangle$ can decay due to the finite $\gamma$ to $|\downarrow \downarrow, 1\rangle$, which destroys the concurrence. Nevertheless, finite $\kappa$ leads to the decay of this state to the initial state $|\downarrow \downarrow, 0\rangle$ and thus closes the cycle. This explains why finite $\kappa$ plays a positive role for the quantum concurrence stabilization. However, if $\kappa$ is too large, $|\uparrow \downarrow, 1\rangle+|\downarrow \uparrow, 1\rangle$ can decay to $|\uparrow \downarrow, 0\rangle+|\downarrow \uparrow, 0\rangle$, which does not allow to attain large enough occupation of $|\uparrow \uparrow, 0\rangle$ via $V_1$. This is why some optimal $\kappa$ is needed to stabilize finite quantum concurrence in the steady state.  At the same time, if the efficiency of $V_2$ is too high, excitation $|\uparrow \downarrow, 1\rangle+|\downarrow \uparrow, 1\rangle$ to the states with higher photon numbers can be too significant, so that small values of $\theta$ do not support finite concurrence in the steady state and some optimal $\theta$ does exist. Thus, finite quantum concurrence appears as a result of a subtle balance between different processes.

We would like to stress that there is no optimal qubit relaxation rate for steady-state concurrence, since energy dissipation in one channel destroys quantum effects in the same channel. As a result, the best qubit relaxation rate for the steady-state quantum concurrence is zero.

An engineering of the photonic 'reservoir' offered by our results might be perspective in the context of quantum information processing. Indeed, a presence of a additional bosonic subsystem with well defined discrete energy levels allows to activate various nontrivial correlations between the qubits by utilizing special types of parametric drivings.

We also analyzed the effect of a pure dephasing. Nonzero pure dephasing rate $\gamma_{\phi}$ tends to destroy the quantum concurrence in the steady state. However, the effect survives provided $\gamma_{\phi} \lesssim \gamma$, although the concurrence becomes smaller. For example, if we assume $\gamma_{\phi} = \gamma$, the steady-state concurrence becomes suppressed by the factor of nearly $1.8$ for the parameters of Fig. \ref{steady}.

Previously, a positive role played by a decay of a bosonic mode which mediates qubit-qubit interaction, was demonstrated for various realizations of hybrid systems mostly under the direct external pumping acting on qubit occupations, which is able to compensate the depopulation of upper levels, see, e.g., Refs. \cite{Kastoryano2011,Krauter2011,Cano2011,Tan2013,Lin2013,Torre2013,Su2014,Joshi2016}. In our system, compensation occurs due to the counterrotating processes leading to the simultaneous qubit excitation and photon creation from vacuum, which makes photon and qubit correlated. However, if modulation becomes too significant (sign-alternating signal), decay of qubit excited states results in an enhancement of quantum effects in photon subsystem such as amplified photon production from vacuum, which suppresses quantum entanglement and correlations within qubit subsystem. Thus, steady-state entanglement appears due to the delicate balance between various processes induced by decoherence and parametric driving.

The entanglement of qubits induced by several parametrical processes was addressed in Refs. \cite{Berman,Felicetti,Solano}. Ref. \cite{Berman} considered a generation of entanglement of two qubits in absence of energy dissipation under the nonadiabatic modulation of cavity frequency leading to the dynamical Casimir effect \cite{VDodonov,ADodonov,X1}. This parametric process is able to create finite quantum concurrence, but it remains weak in the limit of a weak qubit-cavity interaction and in the case of a single switching of a cavity frequency. More sophisticated setup involving two cavities each coupled to its own qubit was considered in Ref. \cite{Felicetti}, see also Ref. \cite{Rossatto}. Quantum correlations of qubits were generated by periodically varying boundary conditions for photon modes between cavities, which led to the quite high qubit-qubit entanglement.

The subject of a very recent paper \cite{Solano} is closest to the present article. There was also considered an entanglement generation between two qubits mediated by a bosonic mode, coupling between two subsystems being modulated not with constant frequency, but with periodically modulated one. Such a modulation signal contains all possible Fourier harmonics, while we argue that the most important role is played by first two harmonics and reveal the effects arising due to the interplay between them, which include an existence of the regime leading to suppression of concurrence but to enhancement of photon generation from vacuum. It was also shown in Ref. \cite{Solano} that cavity decay is able to increase the entanglement, but we additionally demonstrate that there is some optimal decay, so that it should not be too high if we wish to attain maximum steady-state concurrence. As a whole, our results suggest somehow more systematic picture in the subspace of certain parameters, while Ref. \cite{Solano} considers some effects which have not been studied by us at all, such as the dynamics in the case of different modulations of qubits-cavity couplings. Thus, our paper and Ref. \cite{Solano} provide mutually complementary results.

\subsection{Conditional concurrence}

\begin{figure}[h]
\includegraphics[width=0.45\linewidth]{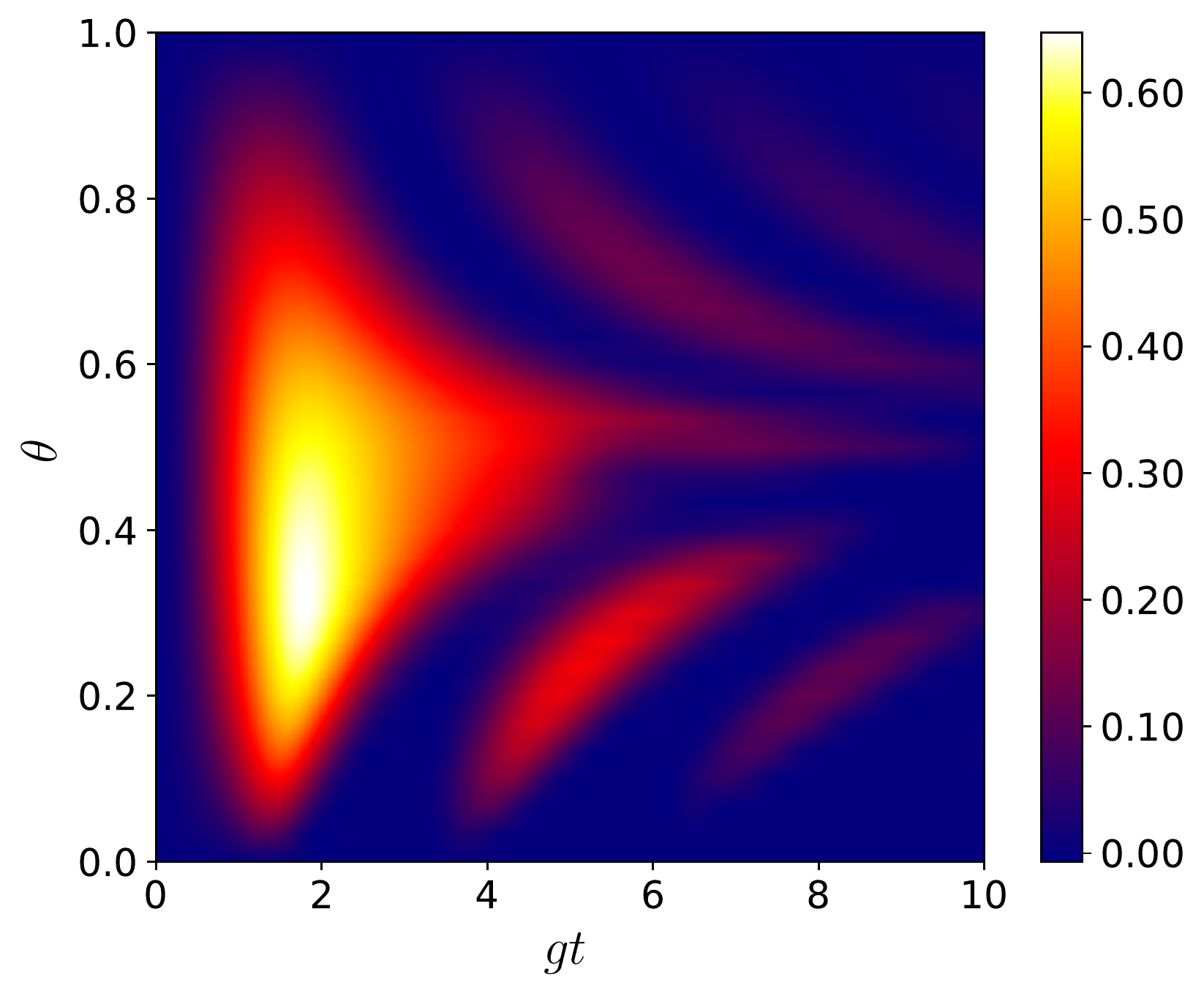}
\caption{
\label{cond}
Color map for the time evolution of the conditional quantum concurrence $C_0$ for zero photon number and different values of parameter $\theta$ at $\kappa=0.001 \omega$, $\gamma=0.01 \omega$, and $g=0.05 \omega$.}
\end{figure}

Conditional concurrences corresponding to different photon numbers can behave in a distinct way compared to the behavior of the total concurrence. For instance, in Fig. \ref{cond} we show an evolution of this quantity for zero photon number. We see that in this case the situation with $\theta < 0.5$ is preferable since it allows to generate a higher concurrence within certain time intervals. Thus, this regime with $\theta < 0.5$ is not useless for the purpose of entanglement engineering, it can be of importance if we desire to design conditional concurrences. It is also evident from this figure that the conditional concurrence is nonzero at $\theta = 0.5$ in a certain time domain, while the total concurrence is identically zero there.

\section{Conclusions}
 \label{summary}

 We studied a dynamics of a two-qubit system coupled to the single-mode quantum resonator at zero temperature under the periodic modulation of a coupling constant between cavity and qubits subsystem. Such a parametric driving leads to activation of counterrotating processes which can be usually neglected in a resonant case and for weakly-interacting and close-to-equilibrium hybrid systems. They are responsible for the parametric qubit excitation and simultaneous photon creation from vacuum.

 Our main finding is the effect of energy dissipation in these parametrically-driven hybrid systems can be highly nontrivial leading to counterintuitive results. Particularly, we have demonstrated that dissipation does not necessarily lead to the suppression of quantum effects, but such effects in one subsystem can be greatly enhanced by finite decay rate in another subsystem. For example, nonzero and optimal cavity decay leads to the stabilization of both the quantum concurrence and quantum mutual information of two qubits. Surprisingly, finite quantum concurrence can survive even in the steady state despite of the finite qubit relaxation rate. In this case, counterrotating processes compensate qubits excited states depopulations by qubit excitation and photon creation. The efficiency of such an excitation is controlled by the balance between two first Fourier harmonics of modulation signal. In the case of strong, i.e., sign-alternating modulation the excitation becomes very efficient, while finite decay rate in qubit subsystem leads to the amplification of photon generation from vacuum. Thus, quantum effects in photon subsystem become enhanced, but the price is the suppression of both the quantum concurrence and quantum mutual information between two qubits in the steady state.

The results we obtained shed a new light on nonstationary cavity QED phenomena. They might be of importance for quantum bath engineering and for the full control on qubits-cavity quantum states. Perhaps, parametrical driving can be used as an alternative or, at least, complementary to error-correction codes approach for maintenance of quantum nature of superconducting circuits.

\begin{acknowledgments}
The authors acknowledge useful comments by E. O. Kiktenko. This work is supported by Advanced Research Foundation (project no. 7/076/2016-2020). W. V. P. acknowledges a support from RFBR (project no. 15-02-02128). D. S. S.  acknowledges a support from Fellowship of the President of Russian Federation for young scientists (fellowship no. SP-2044.2016.5). Yu. E. L. acknowledges a support from RFBR (project no. 17-02-01134).
\end{acknowledgments}

\appendix

\section{Solvable limits and perturbation theory}

It is obvious that if $\theta = 1$, i.e., in the Tavis-Cummings regime, no dynamics is imposed because of the absence of the counterrotating terms. It is however possible to construct a perturbation theory in this case, starting from $\theta = 1$, which is going to indicate a trend as $\theta$ decreases.

We can also consider the situation with $\theta = 0$, which can be referred to as Anti-Tavis-Cummings regime. In this case, only counterrotating terms are present in the Hamiltonian making it exactly solvable in a closed analogy with the usual Tavis-Cummings model (up to the replacement $\sigma_{j,+} \leftrightarrow \sigma_{j,-}$). It is also possible to construct a perturbation theory in this case, as $\theta$ increases.

Surprisingly, it turns out that in the special case $\theta = 0.5$ a solution to the Schr\"{o}dinger equation can be expressed in the compact form. Below we provide all these three solutions and discuss crossovers between them.

\subsection{The vicinity of Tavis-Cummings regime}

The Hamiltonian in the interaction picture within leading order in small parameter $\delta=1-\theta \ll 1$ can be represented as
$$
H=g\theta(V_1+\delta V_2).
$$
The second term can be treated as a perturbation. The time evolution of the wave function is determined by the  Schr\"{o}dinger equation
$$
i\frac{\partial \ket{\Psi(t)}}{g\theta\partial  t}=i\frac{\partial \ket{\Psi(\tau)}}{\partial \tau}=\left(V_1+\delta V_2   \right)\ket{\Psi(\tau)},
$$
where $\tau=g\theta t$. Let us solve this equation by using perturbation theory.

The total wave function can be represented as
$$
\ket{\Psi}=\sum_{n} \left(\alpha_n\ket{\downarrow \downarrow, n}+\beta_{1,n}\ket{\uparrow \downarrow, n}+\beta_{2,n}\ket{\downarrow \uparrow, n}+\gamma_{n}\ket{\uparrow \uparrow, n}\right).
$$
The zero-order solution corresponding to the initial condition $\ket{\Psi^{(0)}(0)}=\ket{\downarrow \downarrow, 0}$ obviously reads as
$$
\ket{\Psi(\tau)}=\ket{\downarrow \downarrow, 0}.
$$
In the first order, it is easy to get equations for amplitudes
$$
i\frac{\partial \alpha^{(1)}_2}{\partial \tau}=\sqrt{2} \left( \beta^{(1)}_{1,1}+\beta^{(1)}_{2,1} \right),
$$
$$
i\frac{\partial \beta^{(1)}_{1, 1}}{\partial \tau}=i\frac{\partial \beta^{(1)}_{2, 1}}{\partial \tau}=\alpha^{(0)}_0+\sqrt{2}  \alpha^{(1)}_{2}+\gamma^{(1)}_{0},
$$
$$
i\frac{\partial \gamma^{(1)}_{0}}{\partial \tau}= \beta^{(1)}_{1,0}+\beta^{(1)}_{2,0}.
$$
The solution is
$$
\alpha^{(1)}_2(\tau)=\frac{\sqrt{2}}{3}\left( \cos (\sqrt{6} \tau) - 1   \right),
$$
$$
\gamma^{(1)}_0(\tau)=\frac{1}{3}\left( \cos (\sqrt{6} \tau) - 1   \right),
$$
$$
\beta^{(1)}_{1,1}(\tau)=\beta^{(1)}_{2,1}(t)=-\frac{i}{\sqrt{6}} \sin (\sqrt{6} \tau).
$$
The concurrence of two qubits in leading order in $(1-\theta)$ is given by
$$
C(t) \simeq \frac{2}{3}(1-\theta) \left( 1-\cos (\sqrt{6} g\theta t)   \right).
$$
It is easy to see that it is produced by a superposition of $\ket{\downarrow \downarrow, 0}$ and $\ket{\uparrow \uparrow, 0}$.

\subsection{The vicinity of Anti-Tavis-Cummings regime}

The Hamiltonian in the interaction picture within leading order in small parameter $\theta \ll 1$ is
$$
H=g(1-\theta)(V_2+\theta V_1).
$$
In order to determine wave function evolution we should solve the Schr\"{o}dinger equation
$$
i\frac{\partial \ket{\Psi(t)}}{g(1-\theta)\partial  t}=i\frac{\partial \ket{\Psi(\tau)}}{\partial  \tau}=\left(V_2+\theta V_1   \right)\ket{\Psi(\tau)},
$$
where $\tau=g(1-\theta)t$.
The zero-order solution for the initial condition $\ket{\Psi^{(0)}(0)}=\ket{\downarrow \downarrow, 0}$ is
$$
\alpha^{(0)}_0(\tau)=\frac{1}{3}\left( 2+ \cos(\sqrt{6}\tau)  \right),
$$
$$
\beta^{(0)}_{1,1}(\tau)=\beta^{(0)}_{2,1}(\tau)=-\frac{i}{\sqrt{6}} \sin(\sqrt{6}\tau),
$$
$$
\gamma^{(0)}_{2}(\tau)=\frac{\sqrt{2}}{3}\left( \cos(\sqrt{6}\tau)  \right).
$$
In this order, the concurrence is identically zero.

In the first order, the equations for amplitudes are
$$
i\frac{\partial \alpha^{(1)}_2}{\partial \tau}=\sqrt{3} \left( \beta^{(1)}_{1,3}+\beta^{(1)}_{2,3} \right)+\sqrt{2} \left( \beta^{(0)}_{1,1}+\beta^{(0)}_{2,1} \right),
$$
$$
i\frac{\partial \beta^{(1)}_{1,3}}{\partial \tau}=i\frac{\partial \beta^{(1)}_{2,3}}{\partial \tau}=\sqrt{3}  \alpha^{(1)}_{2}+2\gamma^{(1)}_{4}+\sqrt{3}\gamma^{(0)}_2,
$$
$$
i\frac{\partial \gamma^{(1)}_4}{\partial \tau}=2 \left( \beta^{(1)}_{1,3}+\beta^{(1)}_{2,3} \right),
$$
$$
i\frac{\partial \gamma^{(1)}_0}{\partial \tau}= \left( \beta^{(0)}_{1,1}+\beta^{(0)}_{2,1} \right).
$$

It is not difficult to find a solution which reads
$$
\gamma_0^{(1)}(\tau)=\frac{1}{3}\left( \cos(\sqrt{6}\tau)-1  \right),
$$
$$
\alpha_2^{(1)}(\tau)=\frac{1}{21\sqrt{2}} \left(9\cos(\sqrt{14}\tau)- 7\cos(\sqrt{6}\tau)-2  \right),
$$
$$
\beta_{1,3}^{(1)}(\tau)=\beta_{2,3}^{(1)}(t)=-\frac{i}{2\sqrt{7}} \left( \sqrt{3}\sin(\sqrt{14}\tau)- \sqrt{7}\sin(\sqrt{6}\tau)  \right),
$$
$$
\gamma_4^{(1)}(\tau)=\frac{\sqrt{6}}{21} \left(3\cos(\sqrt{14}\tau)- 7\cos(\sqrt{6}\tau)  \right).
$$

The resulting concurrence between two qubits in leading order in $\theta$ is given by
$$
C(t) \simeq -\frac{1}{3}\sin^2(\sqrt{6}g(1-\theta)t)+\frac{2}{7}\theta\left(1-\cos(\sqrt{6}g(1-\theta)t) \right)\left(\cos(\sqrt{14}g(1-\theta)t)+\frac{4}{3} \right)
$$
for those values of $t$, which provide positiveness of this quantity, otherwise $C(t)=0$. The first term of this expression is non-positive. The second term competes with the first contribution and it is responsible for finite $C$ at certain time intervals, for which $C(t)>0$. The nonzero $C$ is yielded by a superposition of qubit states $\ket{\downarrow \downarrow}$ and $\ket{\uparrow \uparrow}$.

\subsection*{In the middle between Tavis-Cummings and Anti-Tavis-Cummings regimes}

 We now consider a situation with $\theta=1/2$. The time dependence of the wave function can be determined using the evolution operator
$$
\ket{\Psi(t)}=e^{-iHt}\ket{\downarrow \downarrow, 0}.
$$
The exponents in the right-hand side can be expanded as
$$
e^{-iHt}=1+\sum_{n=1}^{\infty}\left(-it\frac{g}{2}\right)^n\frac{S^n}{n!}\left( a^{\dagger } + a\right)^n,
$$
where $S=\sigma^{+}_1 +\sigma_1 +\sigma^{+}_2+ \sigma_2 $.
The operator $S$ acts only on the qubit sector as
$$
S^n\ket{\downarrow \downarrow}=2^{n-1}
\begin{cases}
\ket{\uparrow \uparrow}+\ket{\downarrow \downarrow},&\text{for even $n$;}\\
\ket{\uparrow \downarrow}+\ket{\downarrow \uparrow},&\text{for odd $n$.}
\end{cases}
$$
Photon creation and annihilation operators only act on photon sector. It can be shown that
$$
\left(a^{\dagger }+a\right)^n\ket{0}=\sum_{k=0}^{n}T(n,k)(a^{\dagger })^k\ket{0},
$$
where
$$
T(n,k)=
\begin{cases}
\dfrac{n!}{2^{\frac{n-k}{2}} \left(\frac{n-k}{2}\right)!k!  },&\text{if $\dfrac{n-k}{2}$ is a non-negative integer;}\\
0,&\text{otherwise.}
\end{cases}
$$

Using these relations, we can represent the wave function as
$$
\ket{\Psi(t)}=\ket{\downarrow \downarrow}\ket{0}+\frac{1}{2}\sum_{n=1}^{\infty}\sum_{k=0}^{n}x^n
\dfrac{2^{\frac{k}{2}}\ket{k}}{\sqrt{k!} \left(\frac{n-k}{2}\right)!   }
\begin{cases}
(\ket{\uparrow \uparrow}+\ket{\downarrow \downarrow}),&\text{for even $n$;}\\
(\ket{\uparrow \downarrow}+\ket{\downarrow \uparrow}),&\text{for odd $n$,}
\end{cases}
$$
where $x=-i\dfrac{g}{\sqrt{2}}t$. After changing an order of summation and summing over $n$ we obtain
\begin{multline}
\nonumber
\ket{\Psi(t)}=\frac{1}{2}\left( 1+e^{x^2} \right)\ket{\downarrow \downarrow}\ket{0}+
\frac{1}{2}\left( e^{x^2}-1 \right)\ket{\uparrow \uparrow}\ket{0}+\\
+\frac{e^{x^2}}{2}\sum_{l=1}^{\infty}\dfrac{2^l x^{2l} }{\sqrt{(2l)!}}
\left(\ket{\uparrow \uparrow}+\ket{\downarrow \downarrow} \right)\ket{2l}+
\frac{x e^{x^2}}{\sqrt{2}}\sum_{l=0}^{\infty}\dfrac{2^l x^{2l} }{\sqrt{(2l+1)!}}
\left(\ket{\uparrow \downarrow}+\ket{\downarrow \uparrow} \right)\ket{2l+1}.
\end{multline}

Now it is possible to recast populations of qubits states $\ket{\downarrow \downarrow}$, $\ket{\uparrow \uparrow}$, $\ket{\downarrow \uparrow}$, $\ket{\uparrow \downarrow}$ in a compact form via corresponding components of density matrix traced over photonic degree of freedom
$$
\rho_{\uparrow \uparrow,\uparrow \uparrow}(t)=\frac{3}{8}+\frac{1}{8}e^{-2g^2 t^2}-\frac{1}{2}e^{-\frac{1}{2}g^2 t^2},
$$
$$
\rho_{\downarrow \downarrow,\downarrow \downarrow}(t)=\frac{3}{8}+\frac{1}{8}e^{-2g^2 t^2}+\frac{1}{2}e^{-\frac{1}{2}g^2 t^2},
$$
$$
\rho_{\uparrow \downarrow,\uparrow \downarrow}(t)=\rho_{\downarrow \uparrow,\downarrow \uparrow}(t)=\frac{1}{8}-\frac{1}{8}e^{-2g^2 t^2}.
$$
It is easy to see that the quantum concurrence of this state is identically zero for any time instance. For the mean photon number, we have a growing in time solution
$$
n_{ph}=\frac{1}{2}g^2 t^2.
$$
These results are in a full agreement with the results of numerical solution of time-dependent Schr\"{o}dinger equation.

Thus, we see that $C$ is identically zero at the ends of the interval $\theta \in [0,1]$ as well as precisely in the middle of it at $\theta = 0.5$. However, we were able to show that $C$ becomes nonzero and maximum $C(t)$ grows as $\theta$ is tuned from the ends of the interval towards $\theta = 0.5$. Thus, there should be two domains, $0<\theta < 0.5$ and $0.5<\theta < 1$, where highest $C(t)$ can be achieved. This is exactly what we see in our numerical simulations, as Fig. \ref{nodecoherence} shows. A comparison between the numerical and analytical approaches near the ends of interval $\theta \in [0,1]$ is presented in Fig. \ref{explic}.


%

\end{document}